\documentclass[twocolumn,showpacs,preprintnumbers,amsmath,amssymb,superscriptaddress,aps,pra,10pt]{revtex4-1}

\usepackage{color}
\usepackage{graphicx}
\usepackage{citesort}

\newcommand{\comment}[1]{}
\newcommand{\ie}{\mbox{i.e.\ }}
\newcommand{\eg}{\mbox{e.g.\ }}

\newcommand{\eqnref}[1]{Eq.~(\ref{#1})}

\newcommand{\figref}[1]{Fig.~\ref{#1}}
\newcommand{\secref}[1]{Section~\ref{#1}}

\newcommand{\eps}{\varepsilon}
\newcommand{\Energy}{{\mathcal{E}}}
\newcommand{\FreeEnergy}{{\mathcal{F}}}
\newcommand{\Power}{{\mathcal{P}}}
\newcommand{\Work}{{\mathcal{W}}}
\newcommand{\total}{{\rm{d}}}
\newcommand{\imag}{{\rm{i}}}
\newcommand{\unitz}{\hat{\rm z}}
\newcommand{\normal}{\hat{\rm n}}
\newcommand{\Lorentzian}{{\rm{L}}}
\newcommand{\Lagrangian}{{\mathcal{L}}}
\newcommand{\Gain}{{G}}
\newcommand{\area}{\mathcal{A}}
\newcommand{\contour}{\mathcal{C}}

\newcommand{\change}{\Delta} 
\newcommand{\var}{\delta\!}
\newcommand{\rot}{\nabla \times}
\newcommand{\cc}{{\rm{c.c.}}}

\renewcommand{\vec}[1]{{\bf{#1}}}
\newcommand{\mytensor}[1]{{\underline{\bf{#1}}}}
\newcommand{\mygreektensor}[1]{{\underline{\boldsymbol{#1}}}}
\newcommand{\dirderiv}{\partial}
\newcommand{\nablaperp}{\nabla_{\!\!\perp}}

\newcommand{\mode}[2]{{#1}^{(#2)}}

\newcommand{\optical}{{\rm{opt}}}
\newcommand{\acoustic}{{\rm{ac}}}

\newcommand{\ePE}{{(\rm{PE})}}
\newcommand{\mPE}{{(\rm{DP})}}
\newcommand{\ES}{{(\rm{ES})}}
\newcommand{\MB}{{(\rm{MB})}}
\newcommand{\opt}{{(\rm{opt})}}
\newcommand{\mech}{{(\rm{mech})}}

\newcommand{\indep}{{\rm{indep}}}
\newcommand{\dep}{{\rm{dep}}}

\newcommand{\acaverage}[1]{\Big \langle \ {#1} \ \Big \rangle_{T_\acoustic}}
\newcommand{\optaverage}[1]{\Big \langle \ {#1} \ \Big \rangle_{T_\optical}}

\begin{document}

\title{Stimulated Brillouin Scattering in integrated photonic waveguides: forces, scattering mechanisms and coupled mode analysis}
\author{C. Wolff}
\affiliation{
  Centre for Ultrahigh bandwidth Devices for Optical Systems (CUDOS), 
}
\affiliation{
  School of Mathematical and Physical Sciences, University of Technology Sydney, NSW 2007, Australia
}

\author{M.~J. Steel}
\affiliation{
  Centre for Ultrahigh bandwidth Devices for Optical Systems (CUDOS), 
}
\affiliation{
  MQ Photonics Research Centre, Department of Physics and Astronomy, Macquarie University Sydney, NSW 2109, Australia
}

\author{B.~J. Eggleton}
\affiliation{
  Centre for Ultrahigh bandwidth Devices for Optical Systems (CUDOS),
}
\affiliation{
  Institute of Photonics and Optical Science (IPOS), School of Physics,
  University of Sydney, NSW 2006, Australia
}

\author{C.~G. Poulton}
\affiliation{
  Centre for Ultrahigh bandwidth Devices for Optical Systems (CUDOS),
}
\affiliation{
  School of Mathematical and Physical Sciences, University of Technology Sydney, NSW 2007, Australia
}

\email{christian.wolff@uts.edu.au}

\date{\today}

\begin{abstract}
  Recent theoretical studies of Stimulated Brillouin Scattering (SBS) in
  nanoscale devices have led to an intense research effort dedicated to the
  demonstration and application of this nonlinearity in on-chip systems.
  The key feature of SBS in integrated photonic waveguides is that small,
  high-contrast waveguides are predicted to experience powerful optical forces
  on the waveguide boundaries,
  which are predicted to further boost the SBS gain that is already expected
  to grow dramatically in such structures because of the higher mode
  confinement alone.
  In all recent treatments, the effect of radiation pressure is included
  separately from the scattering action that the acoustic field exerts
  on the optical field.
  In contrast to this, we show here that the effects of radiation pressure
  and motion of the waveguide boundaries are inextricably linked.
  Central to this insight is a new formulation of the SBS interaction that
  unifies the treatment of light and sound, incorporating all relevant
  interaction mechanisms ---
  radiation pressure, waveguide boundary motion, electrostriction and
  photoelasticity --- from a rigorous thermodynamic perspective.
  Our approach also clarifies important points of ambiguity in the literature,
  such as the nature of edge-effects with regard to electrostriction, and of
  body-forces with respect to radiation pressure.
  This new perspective on Brillouin processes leads to physical insight with
  implications for the design and fabrication of SBS-based nanoscale devices.
\end{abstract}

\maketitle



\section{Introduction}

Stimulated Brillouin Scattering (SBS) is a nonlinear process by which 
light interacts coherently with acoustic vibrations in an optically transparent 
medium~\cite{Boyd2003}. 
Predicted by Brillouin in 1922~\cite{Brillouin1922}, SBS was first 
experimentally demonstrated by Chiao, Townes and Stoicheff soon after the 
invention of the laser~\cite{Chiao1964}, and thereafter became one of the 
standard techniques for measuring the mechanical properties of materials at 
high frequencies~\cite{Uchida1973}. 
SBS has historically often been regarded as a non-desirable side-effect that must be either suppressed or accommodated, 
for example in fiber optics, where it can strongly deplete narrow-band pumps.
However in recent years there has been a remarkable resurgence of interest in 
guided wave SBS~\cite{EggletonBenjaminJ.2013}, driven largely by the ability to harness the 
effect in modern nanophotonics experiments~\cite{Dainese2006,Tomes2009,Grudinin2010,Pant2011,Lee2012}. 
As well as forming the basis for the investigation of fundamental physical effects such as 
slow-light~\cite{Thevenaz2008} and non-reciprocity \cite{Huang2011},
these experiments have led to a number of interesting SBS-based applications, 
such as narrow-linewidth tunable sources~\cite{Abedin2012}, as well as on-chip 
processing of optical~\cite{Pant2012a} and 
radio-frequency~\cite{Vidal2007,Chin2010} signals.
SBS is closely related to the field of opto-mechanics~\cite{Laer2015}, which is
a similarly dynamic aera of research~\cite{Papp2014}.

The SBS interaction arises from a pair of physical mechanisms that transfer 
energy back and forth between the electromagnetic field and the mechanical 
stresses and strains of the material~\cite{Boyd2003,Kobyakov2010}. 
These mechanisms can be categorized as 
{\em scattering} (or forward-action) processes, by which the acoustic modes 
scatter light from one state to another, 
and {\em back-action} (or force-like) processes, 
by which the optical field generates mechanical motion via optical forces and pressures. 
Informed by the understanding of fiber and bulk systems, 
it was thought until very recently that SBS was driven entirely by the 
photoelasticity and electrostriction. 
These are inverse scattering and back-action processes respectively, 
linked by the thermodynamics of dielectric materials under mechanical strain. 
However these mechanisms are not the only possibilities for optomechanical 
interaction: in particular in the field of 
nano-optomechanics it is known that the back-action of radiation pressure can be very large in
small, high-refractive-index-contrast devices, with the effect playing a central role in 
the interaction between optical and acoustic resonators 
in a number of seminal experiments in this 
field\cite{Schliesser2006,Grudinin2009,Lee2012,Eichenfield2009}. 
In recent work Rakich~{\em et al.}~\cite{Rakich2012} showed that radiation pressure 
can also make a significant contribution to the SBS gain: this is the result of
large optical forces that act on the boundaries of suspended waveguides and resonantly excite acoustic modes of 
the free-standing structure. This prediction has important implications for the harnessing of SBS in 
CMOS-compatible materials such as silicon, in which electrostriction is 
relatively weak. Radiation pressure is strongest for small, high index contrast
waveguides, and radiation-pressure-induced SBS has recently been observed in a silicon/silicon nitride 
hybrid waveguide~\cite{Rakich2013}.

In this recent work and in the associated 
literature~\cite{Rakich2012,Rakich2010,Qiu2012}, the scattering of the optical 
mode by the acoustic field is not considered directly; instead the phonon 
generation rate is computed by summing the optical forces due to both radiation 
pressure and electrostriction acting on the waveguide, and the SBS gain is then 
obtained via particle conservation in the classical limit. 
While this is a valid approach, there are several subtleties that arise in the 
physics that are not explicitly discussed in the existing literature. 
The main difficulty with this force-based approach is that force-like terms 
arise in such a way as to make it unclear whether they should be included or 
not. 
In SBS this is most clearly manifested in the form of an electrostrictive 
pressure term that appears on the boundary~\cite{Rakich2013}: it is not entirely
obvious whether this pressure term should be separately included, or whether it 
is simply a manifestation of the radiation pressure and already contained in
the divergence of Maxwell's stress tensor across a material boundary.
In a similar way, the radiation pressure can appear to give rise to
a body force~\cite{Rakich2010,Rakich2012} in addition to the familiar surface 
terms, and it is unclear whether this body force should be included separately 
or whether it is readily contained in the electrostrictive process.
These questions are not easy to answer; indeed the appropriate separation of 
optical forces in materials is related to the proper form of the photon 
momentum within material, and this has been a realm of some debate
(known as the Abraham-Minkowski controversy) for the past hundred years.
Given this uncertainty, the question arises not only as to which forces to 
include in one's calculations, but whether there exist further force-like 
terms that are yet to be discovered.

Our approach to resolve these issues is to avoid optical forces as much as 
possible and found our description on the forward-action processes, which
are not controversial.
These processes include the scattering of light via the photo-elastic 
effect, scattering due to deformations of the waveguide boundary, as well as 
any other process whereby mechanical motion can influence the 
electromagnetic field.
Furthermore, the forward-action half of the SBS interaction is important for 
any experiment in which the stimulated acoustic wave acts on other optical 
fields that are also present in the waveguide, as occurs in the generation 
of frequency combs via SBS~\cite{Buettner2014} or in SBS-based optical 
isolation~\cite{poulton2012design}.
Perhaps more critically, scattering mechanisms form an important part of the 
physics of SBS, and a theory that discusses these explicitly can not only shed 
light on the physicality of back-action processes, but is necessary to 
complete our understanding of Brillouin processes in integrated photonic 
waveguides.

Here we present a new formalism for SBS that considers all interactions in a 
unified way. 
By considering scattering mechanisms explicitly, we derive expressions for the 
interaction that do not rely on the exact form of the optical forces between
and within materials. 
A main result of this formalism is the classification of forward and 
back-action processes into interaction pairs: for example we show that motion 
of the waveguide boundary forms the inverse process to the radiation pressure 
between lossless dielectrics, just as the photoelastic effect is associated 
with the electrostrictive force.
We also identify a new term in the SBS interaction that results from
coupling between the moving dielectric and the magnetic part of the optical
mode.
Furthermore, we observe that optical losses are related to irreversible
forces using the case of radiation pressure as an example; this breaks the
aforementioned symmetry between the scattering and back-action processes.
Finally, we derive coupled mode equations that include all of these effects
in a consistent framework. 
To the best of our knowledge this is the first such treatment in the literature.

The manuscript is structured as follows:
In \secref{sec:preliminaries}, we state the properties of the systems under
investigation and the expansion bases for the optical and acoustic wave
propagation.
In \secref{sec:modal_optical}, we formulate a conventional modal expansion and,
more importantly, state the terms that describe the three scattering mechanisms
of optical modes due to an acoustic displacement field in the waveguide's bulk, 
on its surface and due to by the acoustic velocity field.
These are all first-order terms that are present under our preliminaries.
In \secref{sec:modal_acoustic}, we again formulate a conventional modal 
expansion and then show based on a Lagrangian picture that for reversible 
interactions--and only for those--the coupling constant for the excitation 
of acoustic modes coincides with the previously formulated optical scattering 
constant.
This provides a complete description of SBS that is independent of expressions
for the optical forces and, hence, independent of the Abraham-Minkowski
controversy.
In \secref{sec:discussion} finally, we show how to derive the common resonant 
expression for the SBS-gain of a long waveguide and state limits for its
applicability, we show that the coupled mode model as presented conserves 
energy and finally we discuss how the acousto-optic expressions that arise 
from our treatment are compatible with an expression for the optical
momentum fluxes.
In this context, we also explicitly comment on electrostrictive surface 
pressures.

\section{Preliminaries}
\label{sec:preliminaries}

\begin{figure}
  \centerline{\includegraphics[width=\columnwidth]{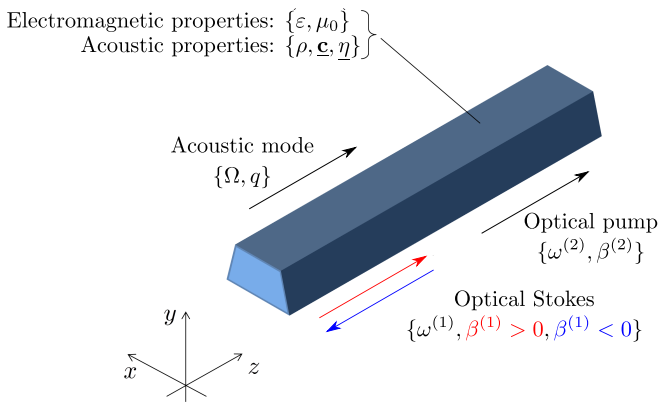}}
  \caption{
    Schematic of the forward (Stokes field represented by the red arrow) and 
    backward (Stokes field in blue) SBS interactions in a waveguide aligned 
    along the $z$-axis. Although a specific waveguide cross-section is shown 
    here, the results apply to waveguides of arbitrary cross section and 
    composition. 
  }
  \label{fig:geometry}
\end{figure}

We consider the interaction between optical and acoustic fields in waveguides 
having the general form depicted in \figref{fig:geometry}, with a material cross 
section that is invariant along the $z$-axis. We make no assumptions on the 
shape of the waveguide, save that it is capable of guiding light in at least 
one optical mode. 
For the materials we assume the absence of magnetic response ($\mu_r = 1$)
and we disregard loss and dispersion in the dielectric parameters. 
We do so mainly to simplify the disucssion in \secref{sec:thermo}; the effect of
dispersion and loss on the optical pulse shape can be easily incorporated in the 
same way as we introduce acoustic loss in \secref{sec:modal_acoustic}.
Furthermore, we neglect the effect of any 
nonlinearity apart from the Brillouin scattering process that we investigate;
specifically, we exclude piezoelectric materials.
The exclusion of material dispersion is justified by the very small frequency 
shifts that occur in SBS, which are typically of only a few GHz.
A material whose permittivity changes appreciably over this frequency range
would probably be too lossy to be used as a material for a waveguide.
If desired, weak dielectric loss can be incorporated in the mode expansion in 
the same way as we incorporate weak acoustic loss 
(see \secref{sec:modal_acoustic}). 
Throughout this work, we use SI units.

\subsection{Electromagnetic part}

We describe the evolution of the optical fields by the electromagnetic wave 
equation in terms of the electric field 
\begin{align}
  \rot \rot \vec E = -\mu_0 \partial_t^2 \vec D; \quad \vec D = \eps \vec E; \quad
  \eps = \eps_r \eps_0.
  \label{eqn:opt_wave_equation}
\end{align}
Here, $\vec E$ and $\vec D$ are the electric field and electric induction field
respectively, and $\partial_t$ denotes the partial derivative with respect 
to time.
In the context of acousto-optics, we prefer the term ``electric induction 
field'' over ``electric displacement field'' to avoid confusion with the 
mechanical displacement field.
Later on, we also use the magnetic field $\vec H$.
The dielectric function $\eps(\vec r)=\eps(x,y)$ is isotropic and homogeneous 
in the \mbox{$z$-direction;} $\vec r = (x,y,z)^T$ is the position vector.
We assume that the electromagnetic fields can be approximated as a 
superposition of two propagating optical eigenmodes 
\begin{align}
  \vec E = & \mode{\vec E}{1} + \mode{\vec E}{2},
\intertext{where}
  \mode{\vec E}{i}(\vec r, t) = & \mode{\vec e}{i}(\vec r, t) \mode{a}{i}(z, t) + \cc\ ,
  \label{eqn:modal_ansatz_opt}
\intertext{and the mode functions factor as}
  \mode{\vec e}{i}(\vec r, t) = & \mode{\widetilde{\vec e}}{i}(x, y) 
  \exp(\imag \mode{\beta}{i} z - \imag \mode{\omega}{i} t),
\end{align}
with frequencies $\mode{\omega}{i}$ and wave vectors 
$\mode{\vec k}{i} = \unitz \mode{\beta}{i}$.
Note that the propagation constants $\mode{\beta}{i}$ may be both positive or 
negative.
The dimensionless envelope functions $\mode{a}{i}$ are assumed to change only 
slowly over an optical wavelength and the time scale of an optical cycle.
The other fields $\vec D$ and $\vec H$ are likewise expanded.
The basis functions $\mode{\widetilde{\vec e}}{i}$ are bound solutions to the
2D-eigenproblem
\begin{align}
  (\nablaperp + \imag \beta \unitz) \times (\nablaperp + \imag \beta \unitz) \times 
  \widetilde{\vec e} = \eps \mu_0 \omega^2 \widetilde{\vec e},
\end{align}
where $\nablaperp$ is the nabla operator in the $x,y$-plane.
There is no need to normalize these modes, although there may be practical 
advantages in doing so in numerical simulations.
Since we neglect dispersion, the average electromagnetic energy density per 
unit length of the waveguide
and the corresponding energy flux carried by the unnormalized mode functions 
are~\cite{Jackson}
\begin{align}
  \mode{\Energy}{i} = & 
  2 \int \total^2 r \ \eps [\mode{\vec e}{i}]^\ast \cdot \mode{\vec e}{i};
  \\
  \mode{\Power}{i} = & 
  2 \int \total^2 r \ \unitz \cdot ([\mode{\vec e}{i}]^\ast \times \mode{\vec h}{i}),
\end{align}
where the integration is across the whole transverse plane.
Using Maxwell's equations, the latter can be recast to an expression that 
only involves the electric field; the energy transport velocity of the mode
(which in the lossless case here is equal to its group velocity)~\cite{Jackson} 
is given by the ratio of the energy flux and energy densities: 
\begin{align}
  \nonumber
  \mode{\Power}{i} = & \frac{1}{- \imag \mode{\omega}{i} \mu_0}
  \int \total^2 r \ \big\{
  [\mode{\vec e}{i}]^\ast \cdot [\unitz \times (\nabla \times \mode{\vec e}{i})] 
  \\
  & \quad +
  [\mode{\vec e}{i}]^\ast \cdot [\nabla \times (\unitz \times \mode{\vec e}{i})]
  \big\}
\label{eq:emflux}
  \\
\label{eq:emvg}
  \mode{v}{i} = & \mode{\Power}{i} / \mode{\Energy}{i}.
\end{align}

\subsection{Acoustic part}

The fundamental equation for the mechanical part of the problem is the acoustic
wave equation~\cite{Auld1990} for the (mechanical) displacement field $\vec U$
\begin{align}
  -\rho \partial_t^2 U_i + \sum_{jkl} \dirderiv_j \left[ c_{ijkl} + \eta_{ijkl} \partial_t \right]
  \dirderiv_k U_l = -F_i,
  \label{eqn:ac_wave_equation}
\end{align}
where $\rho$ is the density, $\mytensor{c}$ is the stiffness tensor and 
$\mygreektensor{\eta}$ is the viscosity tensor.
Here, $\dirderiv_j$ denotes the spatial derivative in the $j$-th spatial direction along $j$,
where \mbox{$j \in \{x,y,z\}$}.
The source term $\vec F$ on the right hand side is the driving external force field per unit volume 
through which the coupling to the electromagnetic field will be introduced.
Assuming first that the acoustic losses are weak, we express the displacement 
field in terms of a solution $\vec u$ with carrier 
$\exp(\imag q z - \imag \Omega t)$ to the lossless wave equation 
(\ie with $\mygreektensor{\eta} = 0$) with a corresponding dimensionless envelope function~$b$:
\begin{align}
  \vec U(\vec r, t) = & \vec u(\vec r, t) b(z, t) + \cc \ ; \\
  \vec u(\vec r, t) = & \widetilde{\vec u}(x, y) \exp(\imag q z - \imag \Omega t); 
\end{align}
where $\widetilde{\vec u}$ is an eigenmode of the equation
\begin{align}
  \rho \Omega^2 \widetilde{u}_i + 
  \sum_{jkl} \ (\nablaperp + \imag q \unitz)_j \ c_{ijkl} 
  \ (\nablaperp + \imag q \unitz)_k \ \widetilde{u}_l = 0.
\end{align}
Note that we need not assume that the acoustic modes are strictly bound, since
significant SBS can occur with leaky acoustic modes~\cite{Poulton2013}. 
However, we assume that the acoustic propagation loss is small so that
it appears to first approximation in the equation of motion for $b$.
The first main advantage of this approach (\ie moving the loss into the dynamics) 
is that the set of functions from which 
we choose $\vec u$ is formed by eigenfunctions to a Hermitian operator resulting 
in convenient orthogonality relations.
Second, $|b|^2$ is related to the amplitude of the acoustic field in a 
straight-forward way throughout the whole system and
so is directly related to the acoustic energy density.
Again, there is no need to normalize the acoustic basis function.
For what follows, the acoustic wave is chosen to be phase-matched with the beat 
between the two optical modes, \ie we specify for the rest of this paper that
\begin{align}
  \Omega = \mode{\omega}{2} - \mode{\omega}{1} \quad \quad 
  \text{and} \quad \quad q = \mode{\beta}{2} - \mode{\beta}{1}.
  \label{eqn:phase_matching} 
\end{align}
Note that such conditions are not in general automatically satisfied if the two
electric fields are chosen freely.  
There must also be an appropriate resonant, phase-matched acoustic mode to 
provide the coupling.
This strict phase-matching condition is used to select appropriate basis
functions for the subsequent modal expansion.

The energy density of the acoustic field is the sum of the kinetic and 
the elastic energy.
For a traveling wave, we focus on the time-averaged total energy
per unit length of the waveguide~\cite{Auld1990}.
For an acoustic mode with unit envelope $b = 1$ it thus reads
\begin{align}
  \Energy_b = & \frac{1}{2} \acaverage{ \int \total^2 r \ 
    \rho |\partial_t \vec U|^2 + \sum_{ijkl} S_{ij} c_{ijkl} S_{kl}
  }\ ,
\end{align}
where $S_{ij} = \frac{1}{2}( \dirderiv_i U_j + \dirderiv_j U_i )$ is the
strain tensor and the subscript $T_\acoustic$ indicates that the average
is taken over a time window that is much longer than one acoustic 
cycle but shorter than the time scale for any relevant slower process.
The transverse integral extends over the interior cross-section
of the waveguide.
If, as is typically the case, the waveguide's total momentum and angular 
momentum are both zero, the average kinetic energy is equal to the average 
elastic energy and we may simplify:
\begin{align}
  \Energy_b 
  = & \acaverage{ \int \total^2 r \ \rho |\partial_t \vec U|^2 }
  = 2 \Omega^2 \int \total^2 r \ \rho |\vec u|^2\ .
\end{align}
The time-averaged energy flux that traverses the waveguide cross section 
$\Power_b$ is given as the normal projection of the product between the 
velocity field and the stress tensor $\mytensor T$~\cite{Auld1990}; the mode's energy transport 
velocity $v_b$ is defined as in \eqnref{eq:emvg}:
\begin{align}
  \Power_b 
  = & - \acaverage{\int \total^2 r \ \unitz \cdot (\partial_t \vec U) \cdot {\mytensor T}}
  \\
  = & - 2 \imag \Omega \int \total^2 r \ \sum_{ikl} c_{zikl} u_i^\ast \partial_k u_l;
  \\
  v_b = & \Power_b / \Energy_b.
\end{align}

Finally, we assume that the coupling mechanism between optical and mechanical modes
is reversible, 
\ie that energy is lost only by the propagation of modes but not by the 
conversion between them.
This means we neglect for example the optical force that occurs due to 
absorption of light.
We comment on this in \secref{sec:conclusion}.

\section{Optical modal equations}
\label{sec:modal_optical}

In this section, we derive the equations of motion for the optical envelope 
functions $\mode{a}{i}$ and examine the two main effects by which a sound wave
can scatter energy from one optical mode into the other.

\subsection{Dynamic equations}

\begin{figure*}
  \includegraphics[width=0.7\textwidth]{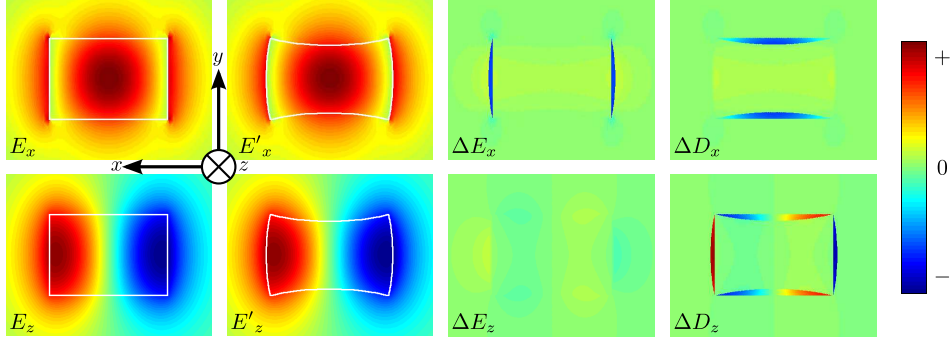}
  \caption{
    The two leftmost columns show the electric field distributions of an
    optical eigenmode in an rectangular waveguide (white outline) before and 
    after a deformation. 
    The other two columns show the deformation-related perturbations of the 
    electric field and the electric induction.
    The waveguide deformation in this figure is grossly exaggerated for 
    reasons of illustration.
    The color scale is linear and symmetric.
  }
  \label{fig:mode_patterns}
\end{figure*}

The mechanical deformation affects the electromagnetic field in two ways.
First, it changes the value of the permittivity.
This is known as the photoelastic effect.
Second, the material boundaries can be displaced and do work on the fields, 
an effect for which no familiar name seems to exist, but which we refer to as
moving boundary scattering.
In either case, the deformation leads to time-dependent changes $\change \vec E$, 
$\change \vec D$
in the electric field and induction, as illustrated by \figref{fig:mode_patterns}.
Furthermore, we allow for an additional magnetization $\change \vec H$ due to
the motion of polarized particles.
In formulating the problem in this way, we go beyond similar prior descriptions
of SBS~\cite{Huang2011}. 

The distorted fields are still solutions of Maxwell's equations and so satisfy 
the wave equation
\begin{align}
  \nonumber
  &
  \rot \rot (\vec E + \change \vec E) 
  + \mu_0 \partial_t^2 (\vec D + \change \vec D) 
  \\
  & \quad
  + \mu_0 \partial_t \rot \change \vec H = 0.
  \label{eqn:perturb_opt_wave_equation}
\end{align}
The field perturbations contain contributions with all possible sum and difference
frequencies, but only those perturbations that simultaneously match the spatial 
and temporal frequency of a basis function (\ie are phase-matched to a basis function)
are relevant for what follows.
Thus, we neglect all but the phase-matched contributions 
\begin{align}
  \nonumber
  \change \vec E(\vec r, t) = & 
  \mode{\change \vec e}{1}(\vec r, t) \mode{a}{2}(z, t) b^\ast(z, t) 
  \\ & \quad
  + \mode{\change \vec e}{2}(\vec r, t) \mode{a}{1}(z, t) b(z, t) + \cc \ ,
\end{align}
Each one of the thus far unspecified patterns $\mode{\change \vec e}{i}$ 
contains the corresponding optical wave carrier:
\begin{align}
  \mode{\change \vec e}{i}(\vec r, t) = &
  \mode{\widetilde{\change \vec e}}{i}(x, y) \ 
  \exp(\imag \mode{\beta}{i} z - \imag \mode{\omega}{i} t),
\end{align}
The other field perturbations $\change \vec D$ and $\change \vec H$ are
treated likewise.
We have explicitly assumed that the field perturbations that are phase-matched to
one optical mode stem from the interaction between the other optical mode and
the sound wave and that they are linear in both fields.
To proceed, we evaluate the contribution from the first mode to 
\eqnref{eqn:perturb_opt_wave_equation}:
\begin{widetext}
\begin{align}
  0 = & \rot \rot (\mode{a}{1} \mode{\vec e}{1} + \mode{a}{2} b^\ast \mode{\change \vec e}{1}) 
  + \mu_0 \partial_t^2 (\mode{a}{1} \mode{\vec d}{1} + \mode{a}{2} b^\ast \mode{\change \vec d}{1})
  + \mu_0 \partial_t \rot (\mode{a}{2} b^\ast \mode{\change \vec h}{1})
  + \cc
  \\
  = &
  \nonumber
  \mode{a}{1} \Big[ \rot \rot \mode{\vec e}{1} + \mu_0 \partial_t^2 \mode{\vec d}{1} \Big]
  + \Big[ \unitz \times (\rot \mode{\vec e}{1})
  + \rot (\unitz \times \mode{\vec e}{1}) \Big] \dirderiv_z \mode{a}{1}
  - 2 \imag \mode{\omega}{1} \mu_0 \mode{\vec d}{1} \partial_t \mode{a}{1}
  \\
  & \quad
  + \mode{a}{2} b^\ast \Big[
    \rot \rot \mode{\change \vec e}{1}  
    + \mu_0 \partial_t^2 \mode{\change \vec d}{1}
    + \mu_0 \partial_t \rot \mode{\change \vec h}{1}
  \Big]
  + \ \text{h.o.t.} + \cc \ ,
\end{align}
\end{widetext}
where `$\text{h.o.t}$' stands for higher order terms in the perturbations and 
higher order derivatives of the envelope functions.
They are neglected since we assumed slowly varying envelopes.
Next, we project onto the mode $\vec e_1$ and average over a time interval 
much longer than the optical time scale.
In this average process, all complex conjugate terms disappear.
Using Eqs.~\eqref{eq:emflux} and~\eqref{eq:emvg}, and dropping all higher order terms, we
obtain
\begin{widetext}
\begin{align}
  \nonumber
  &
  - \imag \mode{\omega}{1} \mu_0 \mode{\Power}{1} \dirderiv_z \mode{a}{1}
  - \imag \mode{\omega}{1} \mu_0 \mode{\Energy}{1} \partial_t \mode{a}{1}
  \\
  = & \mode{a}{2} b^\ast 
  \Bigg\{ \int \total^2 r \quad
    \Big[\mode{\vec e}{1}\Big]^\ast \cdot 
    \Big[ 
      \mu_0 \partial_t^2 \mode{\change \vec d}{1}
      + \rot \rot \mode{\change \vec e}{1}
      + \partial_t \rot \mode{\change \vec h}{1}
    \Big]
  \Bigg\}
  \label{eqn:deriv_opt_1}
  \\
  = & \mode{a}{2} b^\ast \mu_0
  \Bigg\{ 
    \int \total^2 r \quad
    \Big[ \mode{\vec e}{1} \Big]^\ast \cdot 
    \Big[ \partial_t^2 \mode{\change \vec d}{1} \Big]
    - 
    \Big[\partial_t^2 \mode{\vec d}{1}\Big]^\ast \cdot
    \mode{\change \vec e}{1} 
    + 
    \mu_0 \Big[\partial_t \mode{\vec h}{1}\Big]^\ast \cdot
    \partial_t \mode{\change \vec h}{1} 
  \Bigg\}.
  \label{eqn:deriv_opt_2}
\end{align}
\end{widetext}
In the step from \eqnref{eqn:deriv_opt_1} to \eqnref{eqn:deriv_opt_1}, we 
applied two partial integrations (with vanishing boundary terms due to the
modes' exponential localization) to transfer a double curl operation to the
complex conjugate e-field, used the waveguide equation and one of Maxwell's
equations: $\rot \vec E + \partial_t \vec B = 0$.
The second optical mode is treated likewise.
We end up with the final equations for the optical envelope functions, 
where $\mode{v}{1,2}$ are the respective (potentially negative) group 
velocities:
\begin{align}
  \dirderiv_z \mode{a}{1} + \frac{1}{\mode{v}{1}} \partial_t \mode{a}{1}
  = & - \frac{\imag \mode{\omega}{1} \mode{a}{2} b^\ast}{\mode{\Power}{1}}
  Q_1,
  \label{eqn:modal_opt1}
  \\
  \dirderiv_z \mode{a}{2} + \frac{1}{\mode{v}{2}} \partial_t \mode{a}{2}
  = & - \frac{\imag \mode{\omega}{2} \mode{a}{1} b}{\mode{\Power}{2}}
  Q_2.
  \label{eqn:modal_opt2}
\end{align}
Here
\begin{align}
  \nonumber
  Q_1 = & 
    \int \total^2 r \Big[
      [\mode{\vec e}{1}]^\ast \cdot 
      \mode{\change \vec d}{1}
      \\
      & \quad
      - 
      [\mode{\vec d}{1}]^\ast \cdot
      \mode{\change \vec e}{1}
      -
      \mu_0 [\mode{\vec h}{1}]^\ast \cdot
      \mode{\change \vec h}{1}
    \Big],
  \label{eqn:coupling_opt1}
  \\
  \nonumber
  Q_2 = & 
    \int \total^2 r \Big[
      [\mode{\vec e}{2}]^\ast \cdot 
      \mode{\change \vec d}{2}
      \\
      & \quad
      - 
      [\mode{\vec d}{2}]^\ast \cdot
      \mode{\change \vec e}{2}
      -
      \mu_0 [\mode{\vec h}{2}]^\ast \cdot
      \mode{\change \vec h}{2}
    \Big],
  \label{eqn:coupling_opt2}
\end{align}
are works per unit length associated with the couplings.

In a transparent solid insulator, the two effects that lead to the
perturbations $\change \mode{\vec e}{i}$ and $\change \mode{\vec d}{i}$
are the common photoelastic effect and field perturbations caused by 
the changing continuity conditions when a dielectric interface is shifted.
The perturbation $\change \mode{\vec h}{i}$ is caused by an effective dynamic
magnetic coupling effect.
We now discuss these in more detail.

\subsection{Photoelastic effect}
\label{sec:el_photoelasticity}

The term photoelasticity refers to the effect that the electric susceptibility 
of matter changes if it is subject to strain.
In a solid and for small deformations, this can be phenomenologically described
by a fourth rank tensor~\mbox{$\mytensor p$}.
This tensor can be derived from quasi-static or acousto-optic experiments:
\begin{align}
  \chi^\ePE_{ij} = \eps_r^2 \sum_{kl} p_{ijkl} \partial_k U_l,
  \label{eqn:chi_pe}
\end{align}
where we have exploited the 
symmetry $p_{ijkl} = p_{ijlk}$ of the photoelastic tensor.
From \eqnref{eqn:chi_pe} follow the photoelastic parts of the overlap 
integrals in \eqnref{eqn:modal_opt1} and \eqnref{eqn:modal_opt2}:
\begin{align}
  Q^\ePE_1 = &
  \int_A \total^2 r \ [\mode{\vec e}{1}]^\ast \cdot \mode{\change \vec d}{1,\ePE}
  \\
  = &
  \eps_0 \int_A \total^2 r \ \sum_{ijkl} \eps_r^2 
  [\mode{e_i}{1}]^\ast \mode{e_j}{2} p_{ijkl} \partial_k u_l^\ast,
  \label{eqn:coeff_epe}
\end{align}
where $\mode{\change \vec d}{1,\ePE}$ is that part of the physical photoelastic 
polarization field $\change D^\ePE_i = \eps_0 E_j \chi_{ij}^\ePE(\vec U)$ that 
is phase-matched with the optical mode $[\mode{\vec e}{1}]^\ast$.
By interchanging the optical mode labels, we find
\mbox{$Q^\ePE_1 = [Q^\ePE_2]^\ast$}.
Again, the integral is only to be taken over the interior of the waveguide's
material cross section $A$.
Although the displacement field is discontinuous at the waveguide boundary,
the strain field goes from a finite value to zero and the photoelastic
change of the permittivity remains finite.
Boundary effects that are caused by a \emph{displacement} of the material
boundary are treated in \secref{sec:moving_boundary}.

\subsection{Moving polarization effect}
\label{sec:mag_photoelasticity}

It may be surprising that deformation can lead to a magnetic polarization in
a body that was explicitly assumed to have no magnetic susceptibility.
In fact, the absence of magnetic material response only guarantees that a static 
deformation cannot cause such an effect.
A changing mechanical displacement field creates a temporary magnetic 
polarization that is proportional to the electric polarization, because the 
latter describes the dipole moment density of a microscopic charge separation.
When the material is deformed, the separated charges are forced to move and 
form two separated, counter-directed microscopic currents that induce a 
magnetic field at the position of the moving dipole.
This effect is discussed in the context of isolated (electric and magnetic) 
point-dipoles in Ref.~\cite{Hnizdo2012}.
In our case of continuous dipole distributions (\ie polarization fields), the 
effective magnetic polarization is
\begin{align}
  \change \vec H = & (\partial_t \vec U) \times \vec P,
\end{align}
where $\vec P = \vec D - \eps_0 \vec E$ is the total electric polarization.
The phase-matched terms are
\begin{align}
  \mode{\change \vec h}{1} = &
  \imag \Omega \eps_0 (\eps_r - 1) \vec u^\ast \times \mode{\vec e}{2},
\end{align}
and the overlap product with the magnetic induction field $\mu_0 \vec h$ is
after a permutation of the triple product:
\begin{align}
  Q^\mPE_1 = &
  \imag \Omega \mu_0 \eps_0 \int_A \total^2 r \ (\eps_r - 1)  
  \vec u^\ast \cdot (\mode{\vec e}{2} \times [\mode{\vec h}{1}]^\ast).
  \label{eqn:coeff_mpe}
\end{align}
As before, a permutation of mode labels leads to 
\mbox{$Q^\mPE_1 = [Q^\mPE_2]^\ast$}.

This term has not been well-appreciated in the recent literature on SBS.
For optical modes that resemble plane waves, the magnetic induction field
in the overlap integral~\eqnref{eqn:coeff_mpe} can be expressed in terms of 
the optical wave vector and the electric field (see Appendix~\ref{appx:magnetic})
and can be incorporated into the electric photoelastic tensor, where it appears as 
a dispersive, anti-symmetric contribution~\cite{Nelson1971}.
This is the situation \eg in conventional optical fibers but clearly not 
in integrated waveguides such as silicon nanowires.
As a consequence, care must be taken when predicting the SBS-coefficients of 
small waveguides using photoelastic tensor elements that were measured with
high-frequency deformation fields \eg in acousto-optic or SBS experiments.
We will resume this discussion in \secref{sec:forces}.

\subsection{Boundary term}
\label{sec:moving_boundary}

The second important coupling mechanism is caused by the displacement of the
material interfaces of the waveguide as the sound wave propagates along it.
This leads to a strong change in the fields over a very small area exactly 
at the waveguide surface.
This is in contrast to the photoelastic effect, which causes small field 
changes over the full waveguide cross section.
As a consequence, this effect becomes more relevant as the waveguide
cross-sectional area is decreased.

\begin{figure}
  \centering
  \includegraphics[width=\columnwidth]{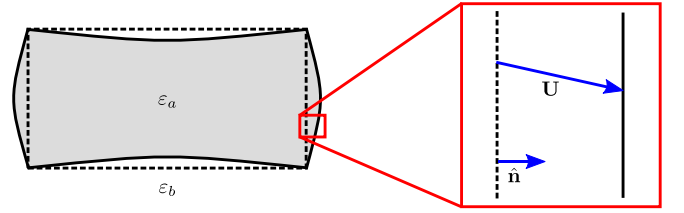}
  \caption{
    Sketch of the boundary displacement effect.
    A waveguide with dielectric permittivity $\eps_a$ is embedded in a 
    background material with permittivity $\eps_b$.
    Please note that the deformation is grossly exaggerated in the left hand
    sketch.
    For surface displacements of realistic magnitude, the quasi-parallel
    shift of the interface by the distance $\normal \cdot \vec U$ is the 
    dominant effect and the second order effects related to a change in the 
    normal vector $\normal$ are negligible.
    This situation is shown in the close-up on the right hand side.
  }
  \label{fig:boundary_effect}
\end{figure}

Clearly, this type of field perturbation appears at every dielectric interface; 
between different solids or liquids and between condensed materials and 
gases or vacuum.
For the sake of illustration, we discuss this using the example of a rectangular
nanowire with permittivity $\eps_a$ surrounded by a domain with another 
permittivity $\eps_b$.
Consider a section of the waveguide outline that is displaced outward as 
illustrated by \figref{fig:boundary_effect}.
We choose the interface normal vector to point outwards.
Maxwell's equations require that the normal component of the induction field
and the in-plane components of the electric fields are continuous across the
interface.
Thus, the electric and induction fields in the space between the old and the
displaced boundary are modified according to
\begin{align}
  \nonumber
  & \vec E^{(\text{before})} = \vec E_{\parallel} + \eps_b^{-1} \eps_0^{-1} \vec D_{\perp} 
  \\
  \longrightarrow & \quad
  \vec E^{(\text{after})} = \vec E_{\parallel} + \eps_a^{-1} \eps_0^{-1} \vec D_{\perp},
  \\
  \nonumber
  & \vec D^{(\text{before})} = \eps_b \eps_0 \vec E_{\parallel} + \vec D_{\perp} 
  \\
  \longrightarrow & \quad
  \vec D^{(\text{after})} = \eps_a \eps_0 \vec E_{\parallel} + \vec D_{\perp},
\end{align}
where the subscripts $\perp$ and $\parallel$ refer to the normal and the 
in-plane parts of the field vectors, respectively.
Thus, we find for the field perturbations in this small area:
\begin{align}
  \change \vec E = & (\eps_b^{-1} - \eps_a^{-1}) \eps_0^{-1}
  \normal (\normal \cdot \vec D), 
  \\
  \change \vec D = & (\eps_a - \eps_b) \eps_0
  (-\normal \times \normal \times \vec E).
\end{align}
Essentially, this is the action of the deformation-related perturbation operator 
on the stress-free solution.
These formulae have already been discussed at length by Johnson \mbox{et al.} 
in the context of perturbation theory for cavity 
eigenfrequencies~\cite{Johnson2002}.

Assuming that the boundary displacement is so small that the field perturbations
are homogeneous in the normal direction, we may replace the integration 
over the whole of the transverse plane 
in Eqs.~\eqref{eqn:modal_opt1} and~\eqref{eqn:modal_opt2}
with a line integration only over all
boundary contours $\contour=\{\contour_i\}$ of the waveguide cross section.
In terms of time-harmonic wave patterns, we find for the coupling coefficients
due to the moving boundary
\begin{align}
  \nonumber
  Q_1^{\MB} 
  = &
  \int_{\contour}  \total \vec r \ 
    (\vec u^\ast \cdot \normal )
    \Big[
    (\eps_a - \eps_b) \eps_0 (\normal \times \mode{\vec e}{1})^\ast
    (\normal \times \mode{\vec e}{2})
    \\
    & \quad 
    - (\eps_b^{-1} - \eps_a^{-1}) \eps_0^{-1} (\normal \cdot \mode{\vec d}{1})^\ast
    (\normal \cdot \mode{\vec d}{2})
  \Big],
  \label{eqn:coeff_mb}
\end{align}
where the factor $(\vec u^\ast \cdot \normal)$ is simply the distance by which the
interface is displaced.
Again, we find by re-labeling optical mode designators that
\mbox{$ Q_2^{\MB} = [Q_1^{\MB}]^\ast $}.
Finally the total coupling in Eqs.~\eqref{eqn:coupling_opt1} 
and~\eqref{eqn:coupling_opt2}
is the sum of the photoelastic and radiation pressure
effects: \mbox{$Q_i = Q_i^{\ePE} + Q_i^{\mPE} + Q_i^{\MB} $}.
Among those, the photoelastic coupling effect is always relevant and dominates
in extended systems such as fibers.
The moving boundary effect can reach comparable magnitude in small high-index
systems, as demonstrated by the Rakich 
group~\cite{Rakich2012,Rakich2013,Rakich2010}.
The moving polarization coupling is usually very weak as we show in 
Appendix~\ref{appx:magnetic}.

\section{Acoustic modal equations}
\label{sec:modal_acoustic}

After our description of the dynamics of the optical envelope functions, we
now turn to the acoustic part of the problem.
First, we derive the dynamic equations and then show how the driving term is
related to the respective driving terms of the optical modal equations already 
found \secref{sec:modal_optical}.

\subsection{Dynamic equations}

We start with \eqnref{eqn:ac_wave_equation}, where the driving force densities $F_i$ are 
due to the electromagnetic fields.
In analogy to the driving terms in the optical wave equation, we assume that the
phase-matched part of the driving force density $\vec F$ is linear in the
two optical envelope functions:
\begin{align}
  \vec F(\vec r, t) = & \vec f(\vec r, t) [\mode{a}{1}(z, t)]^\ast \mode{a}{2}(z, t) + \cc \ .
\end{align}
Substituting this ansatz into equation~\eqref{eqn:ac_wave_equation} and dropping
higher order terms eventually yields
\begin{align}
  \nonumber
  & 
  - \imag \Omega \sum_{jkl} \Big[
  (c_{izkl} \dirderiv_k + \dirderiv_j c_{ijzl}) u_l \dirderiv_z b 
  - 2 \imag \Omega \rho u_i \partial_t b 
  \\
  & \quad 
  + (\dirderiv_j \eta_{ijkl} \dirderiv_k u_l) b
  + [\mode{a}{1}]^\ast \mode{a}{2} f_i \Big] + \cc = 0.
\end{align}
After projecting onto the mode $\vec u$, we end up with the acoustic mode equation
\begin{align}
  \dirderiv_z b + \frac{1}{v_b} \partial_t b + \alpha b = &
  -\frac{\imag \Omega [\mode{a}{1}]^\ast \mode{a}{2}}{\Power_b} Q_b
  \label{eqn:modal_ac}
\end{align}
\begin{align}
  \alpha = & \frac{\Omega^2}{\Power_b} \Bigg\{
    \int \total^2 r \ \sum_{jkl} u_i^\ast \dirderiv_j \eta_{ijkl} \dirderiv_k u_l
  \Bigg\};
  \label{eqn:decay_ac}
  \\
  Q_b = & \int \total^2 r \ \vec u^{\ast} \cdot \vec f,
  \label{eqn:coupling_ac}
\end{align}
where the coupling parameter $Q_b$ is again explicitly a work linear density,
and $1/\alpha$ is the effective dissipation length for the acoustic mode.
This approach to the acoustic part of the SBS process differs significantly from
the treatment in previous works~\cite{Rakich2010,Rakich2012,Rakich2013}.
The main difference is the fact that we regard the sound field as a moving wave rather 
than a localized oscillator.
In \secref{sec:long_wg} we show how to obtain expressions that are
consistent with the literature by assuming that neither the optical nor the
acoustic amplitude vary over the mean acoustic propagation length --- an
approximation that is well justified in long waveguides.
In very short waveguides, however, the propagation of the sound wave can no 
longer be neglected and a treatment like ours is necessary.

Regarding the limitations of our treatment,
it should be stressed that the slowly varying envelope approximation is not 
necessarily justified for the acoustic part of the problem, because
it requires that the acoustic wavelength is much smaller than the length
scale on which the envelope function varies.
This means that $q$ has to be much larger than the damping constant $\alpha$,
and the beat length $\pi/q$ much less than the free propagation length along 
the waveguide.
This is usually the case for backward-SBS and forward-SBS between different
branches of the optical dispersion relation unless they are nearly degenerate.
However, an example for a SBS-setup where the SVEA (and therefore our equations)
is formally not justified can be found in Ref.~\cite{Rakich2013}.
In this work, the waveguide consists of a series of forward-type SBS-active 
suspended regions with a length of $100\,\mu\rm{m}$ each.
This is clearly the maximum free propagation length for the acoustic wave.
The beat length between the optical modes, on the other hand, is given
by the SBS Stokes shift and the optical phase velocity, leading to an
acoustic wavelength of the order of centimeters.
Here, the suspended regions resemble localized harmonic oscillators
and a treatment along the lines of Ref.~\cite{poulton2012design} seems 
appropriate.

\subsection{Optical forces and thermodynamic considerations}
\label{sec:thermo}

We now come to a key part of the analysis, identifying the optical force
density from the optical scattering integral \eqnref{eqn:coupling_opt1}.
There are two common approaches to derive the force that is caused by an optical
field.
The first way is via the Lorentz force.
To this end, the material response is expressed in terms of microscopic charges
and currents which interact with the incident field.
We basically follow this path in \secref{sec:forces}.
The second way is via a thermodynamic potential, see \eg 
Ref.~\cite{Feldman1975} for such a discussion of the connection between 
electrostriction and the photo-elastic effect.
While less familiar in the photonics community, the latter approach is 
attractive for our problem, provided that the change in the entropy is known.

As part of our assumptions, we neglect optical loss and irreversible coupling
effects.
The only source of entropy is the mechanical loss, which can be neglected 
because of the very small acoustic amplitudes in SBS.
If optical loss were to be included, the entropic contribution to the 
thermodynamic potential could become appreciable.
However, the most common experimental situation is a steady state where 
temporally constant optical and acoustic intensities vary spatially along the 
waveguide.
In this case, the temperature would approach an equilibrium distribution that
can be controlled via the properties of a heat sink; the temperature is 
therefore the natural choice for an independent variable.

Typically, the mechanical contribution to the free energy of a solid body 
is separated into boundary terms (surface pressures) and interior
density-like terms (internal stress)~\cite{Landau7}.
However, such a distinction is not convenient for our problem because
of the complexity associated with the moving waveguide boundary.
Accordingly, we adopt a picture based on a displacement field $\vec U$ 
and a driving force density field $\vec F$ that yields both body force 
densities and boundary pressures.
Furthermore, the moving polarization effect depends on 
$\partial_t \vec U$, so a Lagrangian picture is best suited to the problem.
Finally, the electromagnetic continuity conditions force us to decompose
the electric fields in an unusual way.

The variation in the free energy density $\FreeEnergy$ of a waveguide satisfies 
\begin{align}
  \var \FreeEnergy = -S \var T + \var \Energy^\mech + \var \Energy^\opt,
\end{align}
where $\var \Energy^\mech$ and $\var \Energy^\opt$ are the changes in 
mechanical and optical energy, respectively, and $S$ and $T$ denote 
entropy per unit length and temperature. 
The latter are not of great importance in this context as we assume a 
thermodynamically inert process.
We can therefore neglect them.
For the two other terms we assume:
\begin{align}
  \Energy^\mech = & \frac{\rho}{2} |\partial_t \vec U|^2 + \Phi,
  \\
  \Energy^\opt = & \frac{1}{2} \optaverage{\vec E \cdot \vec D +
  \vec H \cdot \vec B },
\end{align}
where $\Phi(\vec U)$ is the stored elastic energy and (recognizing that the
mechanical system cannot follow the rapid electromagnetic oscillations,) the
electromagnetic energy density is averaged over a time window $T_\text{opt}$ that 
includes many optical cycles but is much smaller than one acoustic cycle.
Note that in this context we distinguish between $\vec B$ and
$\mu_0 \vec H$, because we assign the effect of the sound wave to one of them
($\vec H$) while the other one ($\vec B$) is kept fixed as an independent 
variable.
Next, we perform a Legendre transformation to obtain the Lagrangian for 
the opto-mechanical system.
Here, it comes as a great convenience that the electric fields only depend
on $\vec U$ (hence are potential-like) while the magnetic field only depends
on $\partial_t \vec U$ and effectively provides a correction to the kinetic 
energy:
\begin{align}
  \Lagrangian = & \frac{\rho}{2} (\partial_t \vec U)^2 - \Phi
  - \frac{1}{2} \optaverage{\vec E \cdot \vec D -
  \vec B \cdot \vec H }.
\end{align}
The first two terms lead to the acoustic wave equation, whereas the
last two terms correspond to the optical driving force $\vec F$.
By separating these two types of terms in the Euler-Lagrange equations, we
find
\begin{align}
  \vec F = & \rho \partial_t^2 \vec U + \frac{\partial \Phi}{\partial \vec U}
  \\
  \nonumber
  = & - \frac{1}{2} \optaverage{
    \frac{\partial (\vec E \cdot \vec D)}{\partial \vec U} 
  }
  \\
  & \quad
  - \frac{1}{2} \frac{\total}{\total t} \optaverage{
    \vec B \cdot \frac{\partial \vec H}{\partial (\partial_t \vec U)} 
  }\ .
  \label{eqn:opt-force-def}
\end{align}

At this stage it is not yet clear whether the electric part to the 
electromagnetic energy is best expressed with respect to the induction field 
or the electric field as the independent variable. 
The aim must be to formulate the optical field in terms that are not 
influenced by the mechanical displacement field and, thus, allow us to 
describe the optical power independently from the acoustic excitation.
In fact, when the waveguide is deformed as described by $\vec U$, both $\vec E$ 
and $\vec D$ may be perturbed.
A key observation, however, is that there is always a composition of field components 
of $\vec E$ and $\vec D$ that is not changed by the deformation.
This is most easily seen for perturbations of any boundary between different
dielectrics (see \secref{sec:moving_boundary}).
According to the continuity conditions, this unchanged composition consists of 
the normal part of $\vec D$ and the tangential part of $\vec E$;
for the photoelastic effect, which can be described as a change in permittivity 
(see \secref{sec:el_photoelasticity}), the unchanged quantity is $\vec E$.
Such compositions of $\vec E$ and $\vec D$ are independent of the presence of
a weak sound wave and determined only by the choice of the waveguide optical 
modes excited.  
We denote them with a subscript ``$\indep$'':
\begin{align}
  \frac{\partial \vec E_\indep}{\partial U_i} = 
  \frac{\partial \vec D_\indep}{\partial U_i} = & 0.
\end{align}
The dependent variables (subscript ``$\dep$'') then are what remains, \ie the 
difference between perturbed fields and the independent parts:
\begin{align}
  \vec E_\dep = & \vec E + \change \vec E - \vec E_\indep,
  \\
  \vec D_\dep = & \vec D + \change \vec D - \vec D_\indep.
\end{align}
By construction, they completely contain the deformation-dependence of the
optical fields:
\begin{align}
  \frac{\partial \vec E_\dep}{\partial U_i} = 
  \frac{\partial \change \vec E}{\partial U_i}; 
  \quad
  \frac{\partial \vec D_\dep}{\partial U_i} = 
  \frac{\partial \change \vec D}{\partial U_i}.
\end{align}
If we assume that the continuity conditions at material discontinuities 
determine the decomposition of the fields into dependent and independent
quantities, we can furthermore say that
\begin{align}
  \vec E_\indep \cdot \vec D_\indep = \vec E_\dep \cdot \vec D_\dep = 0.
  \label{eqn:field_dep_ortho}
\end{align}
This is also true for electrostriction, because in this case 
\mbox{$\vec D_\indep = \vec E_\dep = 0$}. 
In order to calculate the optical forces, we thus perform another Legendre
transformation:
\begin{align}
  \tilde \Lagrangian = & 
  \Lagrangian - \optaverage{\int \total^2 r \ \vec E_\dep \cdot \vec D}. 
\end{align}
Its electric part satisfies the form
\begin{align}
  \nonumber
  & \optaverage{ \frac{\partial (\vec E \cdot \vec D)}{\partial \vec U} }
  \\
  \nonumber
  = &
  \Big \langle \ 
    \vec E_\indep \cdot \frac{\partial \vec D_\dep}{\partial \vec U}
    - \vec D_\indep \cdot \frac{\partial \vec E_\dep}{\partial \vec U}
    \\
    & \quad
    + \vec E_\indep \cdot \frac{\partial \vec D_\indep}{\partial \vec U}
    - \vec D_\dep \cdot \frac{\partial \vec E_\indep}{\partial \vec U}
  \ \Big \rangle_{T_\optical}
  \\
  = & \optaverage{ \vec E \cdot \frac{\partial \vec D_\dep}{\partial \vec U}
    - \vec D \cdot \frac{\partial \vec E_\dep}{\partial \vec U}
  }\ ,
  \label{eqn:proper_free_energy}
\end{align}
where we used \eqnref{eqn:field_dep_ortho} in the second step.
With this, the optical force density at position $\vec r$ with the 
illumination of the waveguide held constant becomes 
\begin{align}
  \nonumber
  \vec F(\vec r) = &
    \frac{1}{2} \optaverage{
      \vec E 
      \cdot 
      \frac {\partial (\change \vec D)}{\partial \vec U}
      -
      \vec D 
      \cdot 
      \frac{\partial (\change \vec E)}{\partial \vec U}
    }
    \\
    & \quad
    - \frac{1}{2} \frac{\total}{\total t} \optaverage{ \vec B \cdot 
      \frac{\partial (\change \vec H)}{\partial (\partial_t \vec U)} 
    } \ .
  \label{eqn:optical_force}
\end{align}
Next, we note that the total deformation-induced field perturbations 
$\change \vec E$, $\change \vec D$ and $\change \vec H$ are to leading order 
proportional to the displacement field $\vec U$ or its time derivative, 
respectively.
This follows because we evaluate the derivative at the 
point $\vec U = \vec 0$ and any higher order dependence would be to no effect.
Then, the force is independent of the displacement amplitude and the
total work density becomes
\begin{align}
  \Work(\vec U) = & 
  - \int \total^2 r \ \vec U \cdot \vec F
  \\
  \nonumber = & \frac{1}{2} \int \total^2 r \ \sum_{i} U_i
  \optaverage{
    \vec D \cdot \frac{\change \vec E}{U_i}
    -
    \vec E \cdot \frac{\change \vec D}{U_i} 
  }
  \\
  \nonumber & \quad
  + (\partial_t U_i) \optaverage{ \vec B \cdot 
    \frac{\change \vec H}{(\partial_t U_i)} 
  } 
  \\
  & \quad
  - \frac{\total}{\total t} \optaverage{ U_i \vec B \cdot 
    \frac{\change \vec H}{(\partial_t U_i)} 
  } 
  \\
  \nonumber
  = &
  \frac{1}{2} \Big \langle \ \int \total^2 r \ \big[
    \vec D \cdot (\change \vec E)
    -
    \vec E \cdot (\change \vec D) 
    \\
    \nonumber
    & \quad + \vec B \cdot (\change \vec H)
  \big] \ \Big \rangle_{T_\optical}
  \\
  & \quad
  - \frac{\total}{\total t} \optaverage{ \vec U \cdot \left[ \vec B \cdot 
    \frac{\change \vec H}{(\partial_t \vec U) } \right] 
  } \ .
  \label{eqn:work_final}
\end{align}
It is important to point out that the last term oscillates and averages out
to zero over an acoustic cycle.
With this in mind, Eq.~\eqref{eqn:work_final}
is a significant result which confirms that the energy that appears 
as mechanical work per acoustic cycle is precisely the change in the average 
optical energy density.

Using this result, we can evaluate the mechanical overlap integral from 
\eqnref{eqn:modal_ac}.
The coupling integral is supposed to drive the acoustic envelope function 
$b(z, t)$, which we assumed to vary only slowly compared to one acoustic cycle.
So what we really need is a time-average of the mechanical work
\begin{align}
  \acaverage{- \Work(\vec U)} = &
  \acaverage{\int \total^2 r \ \vec U \cdot \vec F} 
  \\
  = &
  a_1^\ast a_2 b^\ast \int \total^2 r \ \vec u^\ast \cdot \vec f 
  \ + \ \cc \ ,
\end{align}
where all terms that oscillate with at acoustic frequencies average to zero.
From \eqnref{eqn:work_final}, we find 
\begin{align}
  \nonumber
  & \acaverage{- \Work(\vec U)} 
  = \frac{a_1^\ast a_2 b^\ast}{2} \int \total^2 r \ \Big\{
  \\
  \nonumber & \quad
  [\mode{\vec e}{1}]^\ast \cdot \change \mode{\vec d}{1}
  + \mode{\vec e}{2} \cdot [\change \mode{\vec d}{2}]^\ast 
  \\
  \nonumber & \quad 
  - [\mode{\vec d}{1}]^\ast \cdot \change \mode{\vec e}{1}
  - \mode{\vec d}{2} \cdot [\change \mode{\vec e}{2}]^\ast 
  \\
  \nonumber & \quad 
  - \mu_0 [\mode{\vec h}{1}]^\ast \cdot \change \mode{\vec h}{1}
  - \mu_0 \mode{\vec h}{2} \cdot [\change \mode{\vec h}{2}]^\ast 
  \\ & \quad
  \Big\} \ + \ \cc \ .
\end{align}
By comparing with Eqs.~(\ref{eqn:coupling_opt1}, \ref{eqn:coupling_opt2},
\ref{eqn:coupling_ac})
and identifying terms with the same dependence on the envelope functions, we find
in the absence of irreversible scattering processes
\begin{align}
  Q_b = \frac{Q_1 + Q_2^\ast}{2}.
  \label{eqn:mech_coupling}
\end{align}
Thus, we have managed to formulate the mechanical driving integral 
\eqnref{eqn:coupling_ac} in terms of the field perturbations, 
and have found a first connection between the scattering strength of the sound 
wave for optical modes and the optical forces that are exerted on the waveguide.

In \secref{sec:modal_optical}, we observed that $Q_1 = Q_2^\ast$, 
which further simplifies \eqnref{eqn:mech_coupling}.
Next, we show that this is not a coincidence but, instead, a consequence
of our initial assumptions and modal approximation.

\subsection{Irreversible forces and optical loss}
\label{sec:inelastic}

Our discussion of forces and scattering effects has certain limitations.
Most importantly, the identities \eqnref{eqn:mech_coupling} and 
\eqnref{eqn:all_coupling} no longer hold for irreversible coupling
mechanisms.
The key feature of such mechanisms is the creation of entropy, \ie the 
connection to loss of some sort.
Although such terms can in principle be generated by the absorption of
phonons (for example, the mechanical friction could separate charges
that contribute to the scattering of light waves), it is likely that the
optical force caused by the absorption of light is most important among 
irreversible coupling effects.

Consider a waveguide composed of an optically lossy material.
The loss may either be due to absorption or to diffuse scattering, \eg 
Rayleigh scattering.
As light is absorbed or diffusely scattered, its momentum is not lost but
transferred to the solid, leading to a longitudinal radiation pressure force,
which can be described as a Lorentz force:
\begin{align}
  \vec F = \optaverage{\vec J \times \vec H} = 
  \sigma \optaverage{\vec E \times \vec H}\ .
\end{align}
Here, both absorption and the time-averaged force are caused by that part of 
the current density $\vec J$ that is in phase with the electric field, \ie that
can be expressed as $\vec J = \sigma \vec E$ with some real-valued conductivity
$\sigma(\omega) = \Im\{\omega \epsilon\}$.
However, this force will not be balanced by a corresponding term in the optical
mode evolution equations, because it is not reversible.
More precisely, the mechanism that causes the force increases the total entropy, 
either by heating the crystal lattice or by inflating the phase space volume of 
the radiation, and thus breaks \eqnref{eqn:work_final}.
The irreversible forces $\vec f^{\text{(irrev.)}}$ result to the addition of the
term \mbox{$Q_b^{\text{(irrev.)}} = \int \total^2 r \ \vec u^{\ast} \cdot 
\vec f^{\text{(irrev.)}}$} to the \eqnref{eqn:mech_coupling}:
\begin{align}
  Q_b = & \frac{Q_1 + Q_2^\ast}{2} + Q_b^{\text{(irrev.)}}.
\end{align}

The occurrence of irreversible contributions is not restricted to radiation
pressure.
It has been reported that the electrostrictive effect of semiconductors 
in a quasi-static electric field is dominated by an irreversible 
process~\cite{Kornreich1967}.
In this case, the finite conductivity leads to both a dissipative current and
a redistribution of charge carriers in reciprocal space that energetically 
favors a distorted crystal lattice.

\section{Discussion}
\label{sec:discussion}

In this and the following section, we discuss some aspects of our model more 
closely.
We do so based on the dynamic equations and coupling terms that we have derived 
up to now:
\begin{align}
  \dirderiv_z \mode{a}{1} + \frac{1}{\mode{v}{1}} \partial_t \mode{a}{1}
  = & 
  - \frac{\imag \mode{\omega}{1} Q_1}{\mode{\Power}{1}} \mode{a}{2} b^\ast,
  \\
  \dirderiv_z \mode{a}{2} + \frac{1}{\mode{v}{2}} \partial_t \mode{a}{2}
  = & 
  - \frac{\imag \mode{\omega}{2} Q_2}{\mode{\Power}{2}} \mode{a}{1} b,
  \\
  \dirderiv_z b + \frac{1}{v_b} \partial_t b + \alpha b 
  = &
  -\frac{\imag \Omega Q_b}{\Power_b} [\mode{a}{1}]^\ast \mode{a}{2},
\end{align}
\begin{align}
  Q_{1} = & Q_{1}^{\ePE} + Q_{1}^{\mPE} + Q_{1}^{\MB}; 
  \\
  Q_{2} = & Q_{2}^{\ePE} + Q_{2}^{\mPE} + Q_{2}^{\MB},
  \\
  Q_b = & \frac{Q_1 + Q_2^\ast}{2} + Q_b^{\text{(irrev.)}}.
\end{align}
Next, we demonstrate which approximations are required to obtain the familiar 
SBS-equations.
Then, we discuss the impact of (approximate) conservation of energy on the
coupling coefficients.
Finally, we will show how the optical force expressions commonly used in the 
literature~\cite{Rakich2012,Rakich2013,Qiu2012} are related to our results and 
briefly comment on the problem of optical forces in general.

\subsection{Gain of long waveguides in steady state} 
\label{sec:long_wg}

We now state the approximations that are required to obtain the well-known 
result for the stimulated Brillouin gain $G$ for steady-state in long waveguides.
In many applications, SBS is a weak process (though still possibly the 
strongest nonlinearity present) and the length scale on which the 
optical power changes is larger than the decay length of the acoustic wave.
Furthermore, SBS is often investigated in a quasi-static setting where all
mode power levels are in equilibrium. 
Thus, the dynamic equations can be approximated for this situation and we obtain 
the expected Lorentzian resonance behavior.
To this end, we need to allow for weak detuning of the laser fields.
This means that the phase-matching conditions \eqnref{eqn:phase_matching} can
no longer both be met at the same time.
However, it is still possible to find modes that fulfill one of them and the
detuning can be expressed either by a frequency difference $\delta \omega$
or a wave vector difference $\kappa = \delta \omega / v_b$.
We assume that the detuning is so small that the eigenmodes, frequencies,
powers and the decay parameter are effectively unchanged.
These assumptions are typically justified except at band edges.
As we are aiming for a steady-state solution, it is advisable to retain the
frequency condition
\mbox{$\Omega = \mode{\omega}{2} - \mode{\omega}{1}$}.
Consequently, we can express the detuning in terms of a wave vector mismatch 
\mbox{$\kappa = q - \mode{\beta}{2} + \mode{\beta}{1}$},
which we incorporate as a slowly varying, harmonic relative phase between
the envelope functions along the waveguides.

First, we impose the steady-state condition, which means that we neglect any
time-derivative in the dynamic equations:
\begin{align}
  \partial_z \mode{a}{1} = &
  - \frac{\imag \mode{\omega}{1} Q_1}{\mode{\Power}{1}} \mode{a}{2} b^\ast, 
  \\
  \partial_z \mode{a}{2} = &
  - \frac{\imag \mode{\omega}{2} Q_2}{\mode{\Power}{2}} \mode{a}{1} b, 
  \\
  \partial_z b + \alpha b= &
  - \frac{\imag \Omega Q_b}{\Power_b} [\mode{a}{1}]^\ast \mode{a}{2}.
\end{align}
Whether for forwards or backwards SBS, the acoustic wave evolves towards positive $z$,
and we solve the last equation by means of its Green's function:
\begin{align}
  \nonumber
  \quad b(z) = & - \frac{\imag \Omega Q_b}{\Power_b} \int_{0}^{\infty} \total z'
  \ \Big\{
  [\mode{a}{1}(z - z')]^\ast 
  \\
  & \quad \times \mode{a}{2}(z - z') \exp(- \alpha z') \Big\}.
\end{align}
Next, we use the assumption that the optical powers vary on a length scale
that is much larger than $\alpha^{-1}$:
\begin{align}
  \nonumber
  & [\mode{a}{1}(z - z')]^\ast \mode{a}{2}(z - z') 
  \\
  \approx & [\mode{a}{1}(z)]^\ast \mode{a}{2}(z) \exp(\imag \kappa z'),
\end{align}
with some detuning parameter $\kappa$ that expresses a violation of the phase 
matching condition.
We find:
\begin{align}
  \nonumber
  b(z) \approx & - \frac{\imag \Omega Q_b}{\Power_b} [\mode{a}{1}(z)]^\ast \mode{a}{2}(z)
  \\
  & \quad \times 
  \int_{0}^{\infty} \total z' \exp[- (\alpha - \imag \kappa) z']
  \\
  = & - \frac{\imag \Omega Q_b}{\Power_b} [\mode{a}{1}(z)]^\ast \mode{a}{2}(z)
  \Lorentzian(\kappa),
\end{align}
where $\Lorentzian(\kappa) = (\alpha - \imag \kappa)^{-1}$ 
is a Lorentzian resonance that defines the bandwidth of the SBS process.
With this and approximating 
\mbox{$\mode{\omega}{1} \approx \mode{\omega}{2} = \omega$}, 
we obtain simplified equations:
\begin{align}
  \partial_z \mode{a}{1} = & \Gain^\ast \mode{\Power}{2} |\mode{a}{2}|^2 \mode{a}{1},
  \label{eqn:long_wg_envelope_gain}
  \\
  \partial_z \mode{a}{2} = & - \Gain \mode{\Power}{1} |\mode{a}{1}|^2 \mode{a}{2},
  \\
  \Gain = & \frac{\omega \Omega Q_1 Q_b^\ast}{
    \mode{\Power}{1} \mode{\Power}{2} \Power_b (\alpha - \imag \kappa)},
\end{align}
with the conventional SBS gain parameter $\Gain$.
In the presence of irreversible force terms, the gain is modified:
\begin{align}
  \Gain = & \frac{\omega \Omega }{\mode{\Power}{1} \mode{\Power}{2} \Power_b (\alpha - \imag \kappa)}
  (|Q_1|^2 + Q_1 [Q_b^{\text{(irrev.)}}]^{\ast}).
\end{align}
The sign of the Stokes mode power $\mode{\Power}{1}$ distinguishes between
propagation in positive and negative $z$-direction, and hence between forward and
backward SBS.
In practice, it is often more convenient to express the light field in terms of
transmitted powers 
\mbox{$\mode{P}{i} = (|\mode{a}{i}|^2 \mode{\Power}{i})$} rather than the 
complex envelope functions themselves.
To this end, we apply the $z$-derivative to the power carried by each mode
and insert \eqnref{eqn:long_wg_envelope_gain}:
\begin{align}
  \partial_z \mode{P}{1}
  = & ( [\mode{a}{1}]^\ast \partial_z \mode{a}{1}
  + \mode{a}{1} \partial_z [\mode{a}{1}]^\ast) \mode{\Power}{1}
  \\
  = &2 \Re\{\Gain\} \mode{P}{1} \mode{P}{2};
  \\
  \partial_z \mode{P}{2}
  = & -2 \Re\{\Gain\} \mode{P}{1} \mode{P}{2},
\end{align}
where, again, negative powers indicate modes propagating in negative 
$z$-direction.
Thus, the SBS-gain relating optical power levels is 
\begin{align}
  \Gamma = & 2 \Re\{\Gain\}
  = \frac{2 \omega \Omega \Re\{Q_1 Q_b^\ast\}}{\mode{\Power}{1} \mode{\Power}{2} \Power_b}
  \cdot \frac{\alpha}{\alpha^2 + \kappa^2}.
\end{align}

\subsection{Conservation of energy}
\label{sec:energy_conservation}

The main finding \eqnref{eqn:work_final} of the previous discussion is not 
restricted to the expansion into guided modes.
It could be possible that the approximation that comes with such a modal 
expansion spoils the conservation laws.
This justifies a check under which conditions the modal equations 
\eqnref{eqn:modal_opt1}, \eqnref{eqn:modal_opt2} and \eqnref{eqn:modal_ac} 
themselves conserve energy.

The total energy that is stored in the optical and acoustic modes is
\begin{align}
  \mathcal{U} = 
  \int_{-\infty}^\infty \total z
  \mode{\Energy}{1} |\mode{a}{1}|^2 +
  \mode{\Energy}{2} |\mode{a}{2}|^2 +
  \Energy_b |b|^2.
\end{align}
Energy is conserved if its time-derivative vanishes:
\begin{widetext}
\begin{align}
  0 = \partial_t \mathcal{U} = &
  \int_{-\infty}^{\infty} \total z \ 
  \partial_t \left[ \mode{\Energy}{1} |\mode{a}{1}|^2 
  + \mode{\Energy}{2} |\mode{a}{2}|^2 + \Energy_b |b|^2 \right]
  \\
  \nonumber
  = & 
  \int_{-\infty}^{\infty} \total z \ \Bigg\{
  \mode{\Energy}{1} [\mode{a}{1}]^\ast \left[
    - \mode{v}{1} \partial_z \mode{a}{1} 
    + \frac{\imag \mode{\omega}{1} Q_1}{\mode{\Energy}{1}} \mode{a}{2} b^\ast
  \right]
  + \mode{\Energy}{2} [\mode{a}{2}]^\ast \left[
    - \mode{v}{2} \partial_z \mode{a}{2} 
    + \frac{\imag \mode{\omega}{2} Q_2}{\mode{\Energy}{2}} \mode{a}{1} b
  \right]
  \\
  & \quad \quad
  + \Energy_b b^\ast \left[ - v_b (\partial_z + \alpha) b
    + \frac{\imag \Omega Q_b}{\Energy_b} [\mode{a}{1}]^\ast \mode{a}{2}
  \right] \Bigg\} \ + \ \cc \ .
\end{align}
\end{widetext}
Clearly, energy can only be conserved in the absence of irreversible forces
and of propagation loss, \ie we have to assume \mbox{$\alpha = 0$} and
\mbox{$Q_b^{\text{(irrev)}} = 0$}.
After collecting terms with identical combinations of envelopes, we find a
condition for the coupling integrals:
\begin{align}
  \mode{\omega}{1} Q_1 - \mode{\omega}{2} Q_2^\ast + \Omega Q_b = 0.
\end{align}
In conjunction with \eqnref{eqn:phase_matching}, this yields
\begin{align}
  & Q_b = \frac{\mode{\omega}{1} Q_1 - \mode{\omega}{2} Q_2^\ast}{\mode{\omega}{1} - \mode{\omega}{2}}
\end{align}
which with \eqnref{eqn:mech_coupling} implies
\begin{align}
  Q_1 = & Q_2^\ast = Q_b \ .
  \label{eqn:all_coupling}
\end{align}
We find that energy can only be conserved if all coupling constants are
equal modulo complex conjugation.
This means in particular that $Q_1 = Q_2^\ast$, an equality that we have
found to be true for the coupling terms in \secref{sec:modal_optical}.
This illustrates that the modal theory is consistent.
At first glance, this conclusion can no longer be drawn for $\alpha \neq 0$.
However, if the coupling mechanism is reversible, loss parameters such as the 
viscosity tensor cannot explicitly appear in the coupling integrals; the 
$Q_i$ can depend on the loss parameters only through the basis functions.
In \secref{sec:preliminaries}, we chose them to be solutions to the lossless
wave equations, because we assumed loss to be so weak that its impact on the
eigenmode pattern can be ignored.
Thus, introducing a non-zero $\alpha$ does not affect the coupling coefficients 
within our approximations and \eqnref{eqn:all_coupling} is valid for weak
propagation loss.
On the other hand, this provides a sanity check to identify situations where
the modal expansion is no longer justified.

\subsection{Remarks on optical forces and the Minkowski-Abraham controversy}
\label{sec:forces}

In the previous section, we showed that the coupling coefficients $Q_i$
for the excitation of the optical and acoustic modes are identical (apart from 
complex conjugation) if energy is conserved in a weak sense, \ie if the coupling
processes conserve entropy and propagation losses are so weak that the 
differences between the actual eigenmodes and those of the lossless wave 
equations are negligible.
In \secref{sec:modal_optical}, we derived expressions for these coupling 
coefficients from the perturbation of the optical eigenmodes caused by the 
deformation of the waveguide.
They include all first-order contributions to the scattering due to boundary 
movement and strain in the material.
As a consequence, we can describe SBS without explicitly referring to optical 
forces.
In this section, we intend to illustrate why this is an advantage.

Momentum $\vec G$ is a conserved quantity and fulfills a continuity equation
\begin{align}
  \partial_t \vec G - \nabla \cdot \mytensor T = 0,
\end{align}
where the columns of the stress tensor $\mytensor T$ play the role of a 
conductive flux for the individual components of the momentum vector.
The temporal change $\vec F = \partial_t \vec G^\mech$ of the mechanical
momentum is called a force.
By decomposing both the momentum and the stress into mechnical 
[superscript $\mech$] and electromagnetic [superscript $\opt$] contributions, 
the optical force naturally appears as a combination of electromagnetic
stress tensor and optical momentum density~\cite{Note1}:
\begin{align}
  \underbrace{\partial_t \vec G^\mech}_\text{total force} = 
    \underbrace{\nabla \cdot \mytensor T^\mech}_\text{mech. force}
    + \underbrace{\nabla \cdot \mytensor T^\opt - \partial_t \vec G^\opt }_\text{optical force}.
  \label{eqn:tot_force}
\end{align}
The term denoted ``mechanical force'' summarizes non-optical force terms, 
\eg gravity, the other two terms comprise the optical force density.
Difficulties arise, because the correct expressions for $\vec G^\opt$ and 
$\mytensor T^\opt$ are subject to a long standing controversy within physics.
Apart from the two best known forms of these quantities by Abraham and 
Minkowski, numerous further expression (some for special cases such as 
non-magnetic media or static fields) have been proposed.
A slightly dated but very informational review of this controversy was written
by Brevik~\cite{Brevik1979}.
Brevik discusses the compatibility of different forms of $\mytensor T^\opt$ 
and $\vec G^\opt$ with several experiments and finds that the question for
the best-suited expressions may depend on the exact physical situation, \eg 
whether the fields are static or dynamic or if some effects (magnetostriction, 
electrostriction or radiation pressure) can be neglected.
The experiments that Brevik reviewed usually only capture certain aspects of 
the total optical force while neglecting others (a situation similar to SBS 
in optical fibres, where only electrostriction contributes).
It is not obvious from the outset that electrostriction and radiation pressure
are complementary and exactly add up to the total optical force.
However, the typical expressions used for SBS in nanowires (including both 
electrostriction and radiation pressure~\cite{Rakich2012,Rakich2013,Qiu2012}) 
could in fact double-count contributions to the true coupling.
A natural candidate for this would be the electrostrictive boundary pressure 
term.
Intuitively it is not clear that this term is completely independent of the 
expression for the radiation pressure.
The other question is whether further interaction terms could have been missed 
in the literature.
The easiest way to answer these questions is to avoid explicitly stating optical 
forces and to approach the problem from the electromagnetic side; the route 
we chose for the first part of this paper.

We can now reinterpret our coupling terms from
Secs.~\ref{sec:el_photoelasticity}--\ref{sec:moving_boundary}
as overlap products of the form 
\begin{align}
  Q_b & = \int \total^2 r \ \vec u^{\ast} \cdot \vec f
  \label{eqn:qb_opt_force_form}
\end{align}
We find for the force associated with the moving boundary 
(\eqnref{eqn:coeff_mb}):
\begin{align}
  \vec f^\MB & = \normal 
  \Big[
    (\eps_a - \eps_b) \eps_0 (\normal \times \mode{\vec e}{1})^\ast
    (\normal \times \mode{\vec e}{2})
    \\
    & \quad 
    - (\eps_b^{-1} - \eps_a^{-1}) \eps_0^{-1} (\normal \cdot \mode{\vec d}{1})^\ast
    (\normal \cdot \mode{\vec d}{2})
  \Big],
\end{align}
where $\normal$ is the local normal vector of the interface.
This is the familiar expression for the radiation pressure on the waveguide 
boundary.
The only difference is that in the literature this is derived from Maxwell's
stress tensor and, hence, only justified for interfaces between a dielectric
and vacuum.
The surface pressure between two dielectrics would have to be derived from the
correctly generalized stress tensor in matter.
In contrast, our derivation only relies on the continuity conditions derived
from Maxwell's equations and is valid for any combination of materials.
The advantage of our derivation is the confidence given that the familiar 
expression for the radiation pressure is correct in every situation.

Next, we find for the force associated with the moving polarization effect
(\eqnref{eqn:coeff_mpe}):
\begin{align}
  \vec f^\mPE & = \imag \Omega 
  (\mode{\vec d}{2} - \eps_0 \mode{\vec e}{2}) \times [\mode{\vec b}{1}]^\ast.
    \label{eqn:mpe_body_force}
\end{align}
The cross product is the difference between the Minkowski momentum and the 
Abraham momentum for light inside the material. 
In the context of Nelson's work~\cite{Nelson1991}, it appears as the 
pseudomomentum flux carried by optical phonons inside the material, \ie that
part of the optical momentum that is tied to the dielectric's frame of 
reference.
The term \eqnref{eqn:mpe_body_force} describes the advective momentum transport
inside the material and has been missed in the literature.
However, we have already argued that it is very small and very likely to be of 
no importance in the context of SBS.

Finally, the overlap integral of the electric photoelastic effect 
\eqnref{eqn:coeff_epe} cannot trivially be rewritten in the form of
\eqnref{eqn:qb_opt_force_form}.
This is because this optical force (electrostriction) naturally arises as a
material stress.
However, in practice the waveguide usually consists of domains $\area^{(\nu)}$
composed of materials with constant or continuously varying parameters $\eps_r^{(\nu)}$
and $p_{ijkl}^{(\nu)}$, where the superscript $(\nu)$ is a domain index.
We now apply the divergence theorem individually to each domain to find:
\begin{align}
  \nonumber
  & \eps_0 \int_{\area^{(\nu)}} \total^2 r \ \sum_{ijkl} [\eps^{(\nu)}_r]^2 
  [\mode{e_i}{1}]^\ast \mode{e_j}{2} p^{(\nu)}_{ijkl} \partial_k u_l^\ast
  \label{eqn:epe_stress}
  \\
  \nonumber
  & = 
  \eps_0 \sum_{ijkl} 
  \int_{\contour^{(\alpha)}} \total^2 r \ u_l^\ast n_k 
  [\eps^{(\nu)}_r]^2 [\mode{e_i}{1}]^\ast \mode{e_j}{2} p^{(\nu)}_{ijkl}
  \\
  & \quad - 
  \eps_0 \sum_{ijkl} \int_{\area^{(\nu)}} \total^2 r \ u_l^\ast 
  \partial_k \left( [\eps^{(\nu)}_r]^2 [\mode{e_i}{1}]^\ast \mode{e_j}{2} 
  p^{(\nu)}_{ijkl}\right),
\end{align}
where $\contour^{(\nu)}$ refers to the contour that surrounds 
$\area^{(\nu)})$ and $\normal$ is the local normal vector.
From this expression, we can now extract two contributions to the 
electrostrictive force: a body force inside the $\nu$th domain
\begin{align}
  f^{(\text{ePE, body}, \nu)}_l = & - \eps_0 \sum_{ijkl} 
  \partial_k \left([\eps^{(\nu)}]_r^2 [\mode{e_i}{1}]^\ast \mode{e_j}{2} 
  p^{(\nu)}_{ijkl}\right),
  \label{eqn:epe_body_force}
\end{align}
and a pressure on the boundary surrounding the $\nu$th domain
\begin{align}
  f^{(\text{ePE, boundary}, \nu)}_l = & 
  \eps_0 \sum_{ijkl} n_k [\eps^{(\nu)}_r]^2 [\mode{e_i}{1}]^\ast 
  \mode{e_j}{2} p^{(\nu)}_{ijkl}.
\end{align}
These terms are the divergence of the electrostrictive stress tensor and its
normal projection to an interface.
They are the common expressions for electrostrictive coupling in the literature.
From this, we can conclude that the electrostrictive boundary pressure is
a consequence of formulating the electrostrictive coupling using a force density
rather than the more natural stress.
As such it has no physical meaning and is not related to radiation pressure.
It is entirely a matter of taste and convenience whether to prefer the overlap 
of a stress field and a strain field or separate overlap of a displacement 
field with a body force and a boundary pressure.
Furthermore, \eqnref{eqn:mpe_body_force} and \eqnref{eqn:epe_body_force} are
the only body forces inside a lossless dielectric.
It is not surprising that this is the expression for electrostrictive coupling,
because photoelasticity and electrostriction are connected via a Maxwell
relation.

\section{Conclusion}
\label{sec:conclusion}

In this paper, we have formulated the dynamics of SBS in a coupled mode 
framework.
Based on energy considerations, we have established connections between the 
nonlinear coupling coefficients that mediate the interplay of optical and
acoustic eigenmodes in a waveguide.
In this context, the connection between scattering of optical eigenmodes due to
the motion of dielectric interfaces in conjunction with the electromagnetic
continuity conditions on the one hand and the transverse radiation pressure
on the other hand is a finding of some interest.
We have also pointed out that irreversible forces are bound to appear in lossy 
waveguides and that these coupling terms form a qualitatively different
contribution to the SBS gain.
This finding may become relevant in the near future as people start to begin
to consider metals in SBS-designs in order to further reduce the device 
dimensions.
Finally, we have shown how the miniaturization is challenged by the finite
velocity of acoustic waves.
Again, this finding is of certain importance for the design of integrated
SBS-devices.

\section*{Acknowledgments}
\noindent
We are deeply indebted to Gustavo Wiederhecker for fruitful discussions.
Furthermore, we acknowledge financial support of the Australian Research 
Council via Discovery Grant DP130100832,
its Laureate Fellowship
(Prof. Eggleton, FL120100029) program and the ARC Center of Excellence CUDOS
(CE110001018). 
\vspace{5mm}

\begin{appendix}

\section{Estimate of moving polarization effect for plane waves}
\label{appx:magnetic}

In this appendix, we estimate the relevance of the magnetic coupling term 
described in \secref{sec:mag_photoelasticity}.
We base this on the expressions for the reciprocal optical force contributions
as identified in \secref{sec:forces}.
The contribution from the magnetic coupling depends heavily on details of the
structure, especially on the geometry and mode symmetry.
Thus, a discussion of its influence on the SBS-properties of specific designs 
for integrated optical waveguides is beyond the scope of this paper.
On the other hand, the problem of plane waves can be easily solved and is
important to estimate the contribution of the magnetic part in measurements
of the photoelastic parameters of a material based on acousto-optic methods.

As we assume quasi-plane waves, we are dealing only with the bulk material, 
which we furthermore assume to be isotropic.
Finally, we assume backward-type SBS caused by the longitudinal acoustic
wave.
The force caused by the reciprocal process of the conventional electric 
photoelastic effect and by the reciprocal process of the dynamic, effective
magnetic coupling effect are:
\begin{align}
  \vec F^\ePE = & \nabla \cdot \mytensor T^\ES ,
  \\
  \vec F^\mPE = & \partial_t (\vec P \times \vec B).
\end{align}
Both terms oscillate at optical frequencies. 
The appropriate time-averages are:
\begin{align}
  \optaverage{\vec F^\ePE} = & \partial_z \left[ \eps_0 \eps_r^2 p_{xxzz} |\vec E|^2 \right]
  \\
  = & -\imag \Omega \frac{\eps_0 \eps_r^2}{c_\acoustic} p_{xxzz} |\vec E|^2,
  \\
  \optaverage{\vec F^\mPE} = & - \imag \Omega (\vec P \times \vec B) 
  \\
  = & -\imag \Omega \frac{\eps_0 (\eps_r - 1) \sqrt{\eps_r} }{c} |\vec E|^2,
\end{align}
where $c$ and $c_\acoustic$ are the vacuum speed of light and the speed of the
longitudinal acoustic wave, respectively.
The ratio between both forces is
\begin{align}
  \frac{\optaverage{\vec F^\mPE}}{\optaverage{\vec F^\ePE}}
  = & \frac{c_\acoustic}{c} \cdot \frac{(\eps_r - 1)}{\eps_r^{3/2} p_{xxzz}}.
\end{align}
The first factor is proportional to \mbox{$\Omega / \omega$}, the second factor
is of the order of $1$.
This means that the magnetic coupling is a very weak effect for plane waves.

\end{appendix}


\end{document}